\documentclass{iau} 
\usepackage{graphicx}

\def\aap{\textit{A\&A}}
\def\mnras{\textit{MNRAS}}
\def\aj{\textit{AJ}}
\def\apj{\textit{ApJ}}
\def\apjs{\textit{ApJS}}
\def\apjl{\textit{ApJ Letters}}
\def\nat{\textit{Nature}}
\def\sun{$_{\odot}$\ }

\title[Intragroup and Intracluster Light] 
{Intragroup and Intracluster Light}

\author[Chris Mihos]   
{J. Christopher Mihos$^1$}

\affiliation{$^1$Department of Astronomy, Case Western
Reserve University}

\pubyear{2015}
\volume{317}  
\jname{The General Assembly of Galaxy Halos}
\editors{A.Bragaglia, M. Arnaboldi, M. Rejkuba, \& D. Romano, eds.}
\begin{document}

\maketitle

\begin{abstract}
The largest stellar halos in the universe are found in massive galaxy
clusters, where interactions and mergers of galaxies, along with the
cluster tidal field, all act to strip stars from their host galaxies and
feed the diffuse intracluster light (ICL) and extended halos of
brightest cluster galaxies (BCGs). Studies of the nearby Virgo Cluster
reveal a variety of accretion signatures imprinted in the morphology and
stellar populations of its ICL. While simulations suggest the ICL should
grow with time, attempts to track this evolution across clusters
spanning a range of mass and redshift have proved difficult due to a
variety of observational and definitional issues. Meanwhile, studies of
nearby galaxy groups reveal the earliest stages of ICL formation: the
extremely diffuse tidal streams formed during interactions in the group
environment.

\end{abstract}

\firstsection 
\section{Introduction}

In considering the connection between accretion and the formation of
galaxy halos, perhaps nowhere is the process more dramatically
illustrated than in the assembly of the most massive halos -- the
extended BCG envelopes and diffuse intracluster light (ICL) that is
found in the centers of massive galaxy clusters. Unlike quiescent field
galaxies whose major accretion era lies largely in the past, under
hierarchical accretion scenarios, clusters of galaxies are the most
recent objects to form (\eg Fakhouri \etal\ 2010); their massive central
galaxies continue to undergo active assembly and halo growth even at the
current epoch, and may have accreted as much as half their mass since a
redshift of $z=0.5$ (\eg de Lucia \& Blaizot 2007). Thus the cluster
environment presents an ideal locale for studying the accretion-driven growth of massive
galaxy halos.

\begin{figure}[b]
\begin{center}
 \includegraphics[width=5.25in]{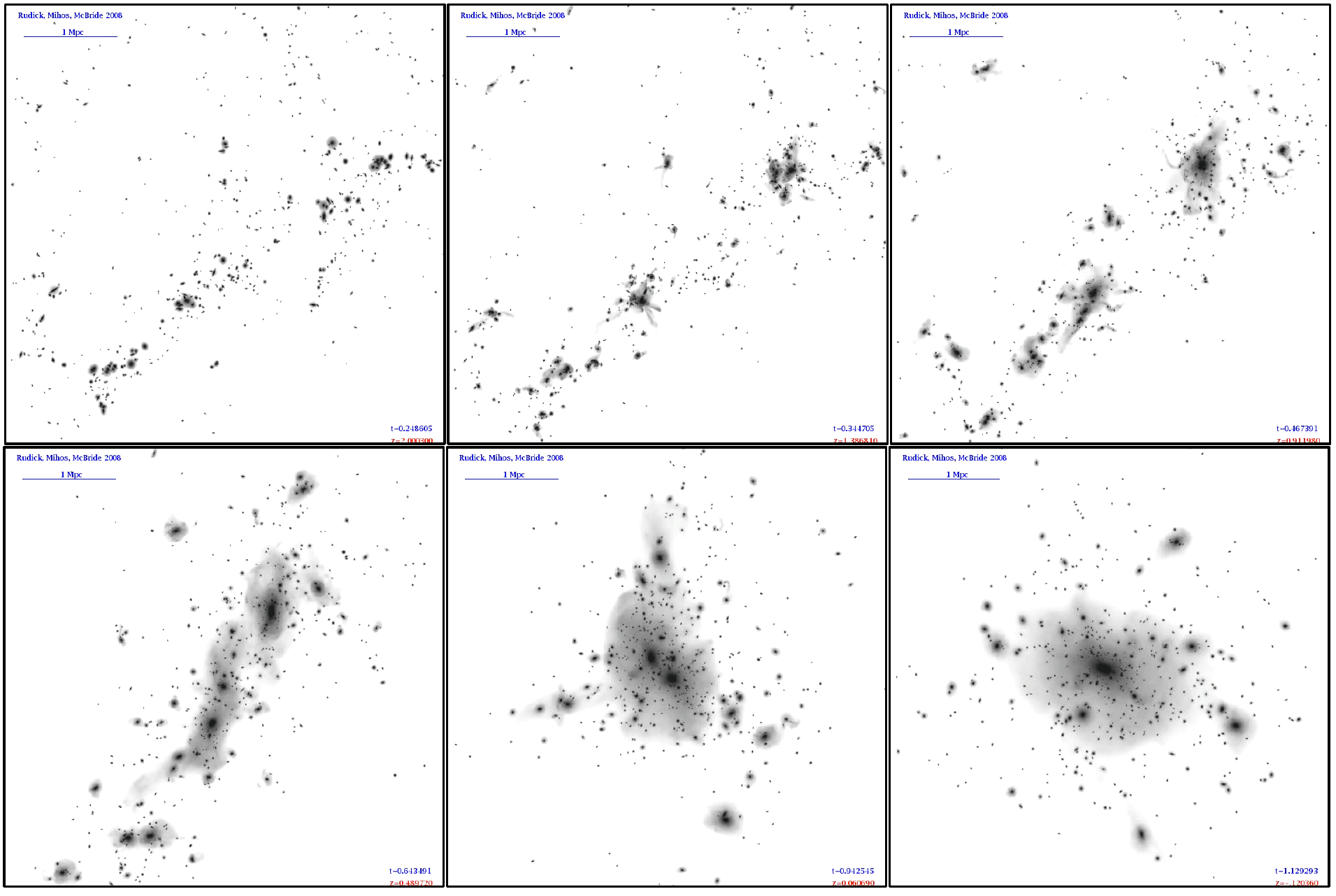}
 \caption{Formation of intracluster light during the assembly of a $10^{15} M_{\odot}$
 galaxy cluster. The panels run from $z = 2$ (upper left) to the present day (lower right). 
 From Rudick \etal\ (2011).}
   \label{sim}
\end{center}
\end{figure}

As galaxy clusters assemble, their constituent galaxies interact with
one another, first within infalling groups, then inside the cluster
environment itself. Over the course of time, a variety of dynamical
processes liberate stars from their host galaxies, forming and feeding
the growing population of intracluster stars. This complex accretion
history is illustrated in Figure~\ref{sim}, using the collisionless
simulations of Rudick \etal\ (2011). At early times,
individual galaxies are strewn along a collapsing filament of the cosmic
web. Gravity quickly draws these galaxies into small groups, which then
fall together to form larger groups. In the group
environment, slow interactions between galaxies lead to strong tidal
stripping and the formation of discrete tidal tails and streams. As the
groups fall into the cluster, this material is efficiently mixed into the cluster
ICL (Rudick \etal\ 2006, 2009). Concurrently, mergers of
galaxies in the cluster core expel more stars into intracluster space
(Murante \etal\ 2007), as does ongoing stripping of infalling
galaxies due to interactions both with other cluster galaxies and with
the cluster potential itself (Conroy \etal\ 2007, Purcell \etal\
2007, Contini \etal\ 2014). Additionally, even {\it in-situ} star
formation in the intracluster medium, from gas stripped from infalling
galaxies, may contribute some fraction of the ICL as well (Puchwein
\etal\ 2010). All these processes lead to a continual growth of the
intracluster light over time, as clusters continue to be fed by
infalling groups and major cluster mergers. This evolution predicts that
 ICL properties should be linked to the dynamical state of the
cluster -- early in their formation history, clusters should be marked
by a low {\it total} ICL fraction but with a high proportion of light in
cold (and more easily visible tidal streams), while more evolved
clusters would have higher ICL fractions found largely in a smooth,
diffuse, and well-mixed state.

The fact that these various processes all operate concurrently makes it
difficult to isolate their individual contributions to the ICL, and
computational studies differ on whether group accretion, major mergers,
or tidal stripping dominate the ICL. Fortunately, these processes
imprint a variety of observable signatures in the ICL. The morphology
and color of the diffuse light as well as the spatial distribution and
kinematics of discrete ICL tracers (red giant branch (RGB) stars,
planetary nebulae (PNe), and globular clusters (GCs)) all have potential
to disentangle the ICL formation channels. For example, the galaxy
mass-metallicity relationship predicts that stripping of low mass
satellites would deposit preferentially metal-poor stars into the ICL,
while mergers of massive galaxies would lead to more metal-rich ICL.
Similarly the age distribution of ICL populations may differentiate
between stripping of old stellar systems versus that from star-forming
galaxies, or even contributions from {\it in-situ} ICL production. Thus,
observational studies of the morphology, colors, kinematics, and stellar
populations in the ICL are well-motivated to track the detailed
accretion histories of massive clusters.

\section{ICL in the Virgo Cluster}

\begin{figure}[b]
\begin{center}
 \includegraphics[width=5.25in]{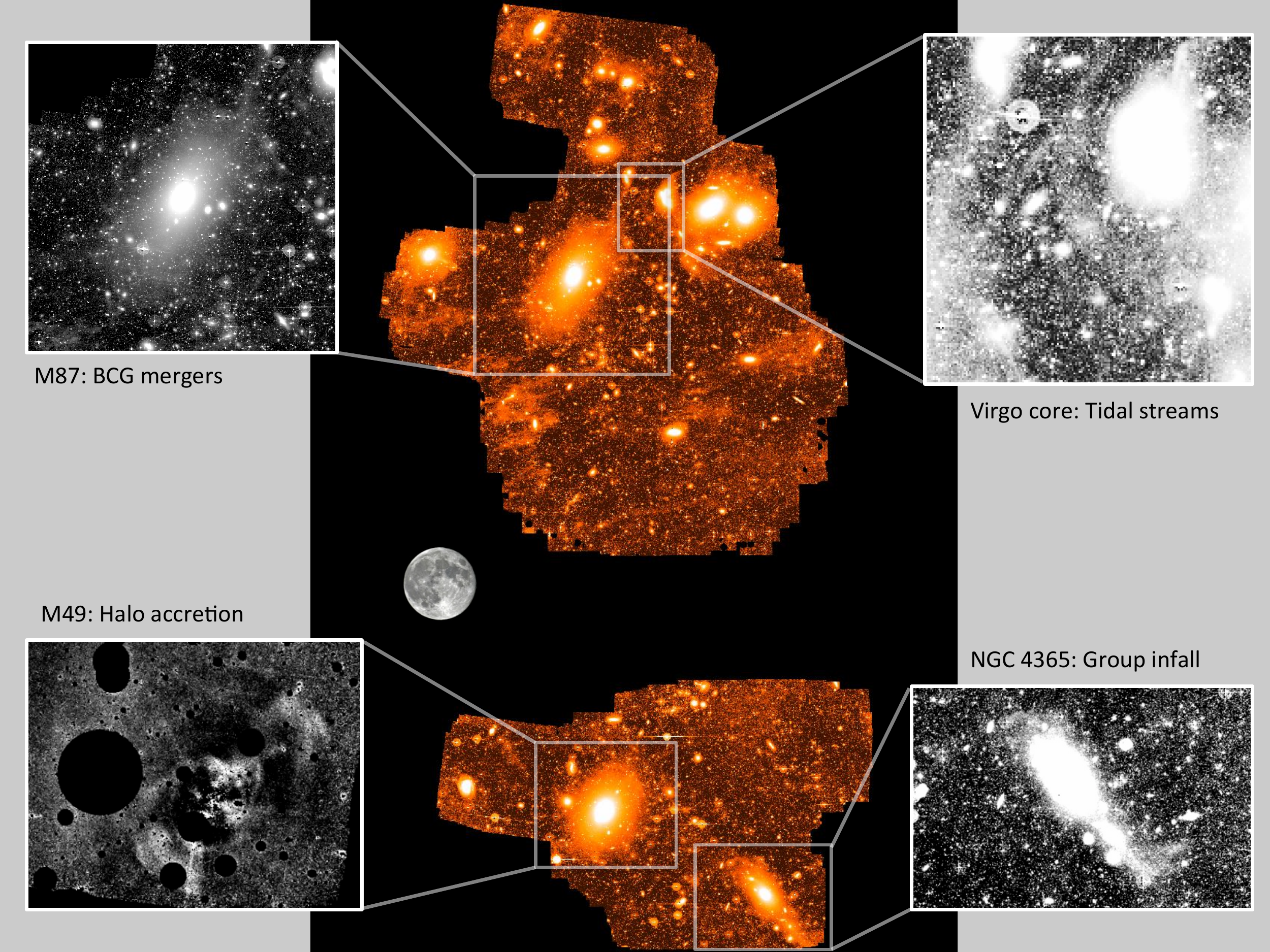}
\caption{Diffuse light in the Virgo Cluster. The center panel shows deep
($\mu_{V,lim} \approx 28.5$) wide-field imaging of Virgo taken using
CWRU's Burrell Schmidt telescope (Mihos \etal\ in prep); the inset moon
shows a 30$^\prime$ scale. Panels show M87's extended halo (upper left;
Mihos \etal\ 2005), tidal streams in the Virgo core (upper right; Mihos
\etal\ 2005, 2015), M49's system of accretion shells (lower left;
Janowiecki \etal\ 2010, Mihos \etal\ 2013) and the diffuse intragroup
light surrounding NGC 4365 (lower right; Bogd\'an \etal\ 2012, Mihos
\etal\ in prep).}
   \label{virgo}
\end{center}
\end{figure}

We can use the nearby Virgo Cluster to illustrate the wealth of
information locked in the ICL. Figure~\ref{virgo} shows deep, wide-field
imaging of the Virgo Cluster taken using CWRU Astronomy's Burrell
Schmidt telescope (Mihos \etal\ in prep). Covering 16 square degrees
down to a surface brightness of $\mu_V \sim 28.5$, the imaging reveals
the complex web of diffuse light spread throughout the core of Virgo. A
number of tidal streams are visible, most notably two long ($>$ 100 kpc)
thin streams NW of M87 (Mihos \etal\ 2005, Rudick \etal\ 2010). Smaller
streams are also found around the Virgo ellipticals M86 and M84, likely
due to stripping of low mass satellite galaxies, as well as a system of
shells and plumes around M89 suggestive of one or more major mergers
(Malin 1979, Janowiecki \etal\ 2010). However, the total luminosity
contained in these discrete streams is only $\sim 1-2\times
10^9$~L\sun\llap; the bulk of the ICL is likely found in more diffuse
form, locked in the extended halo of M87 or strewn throughout the
cluster at lower surface brightness.

Indeed the deep imaging reveals not only the thin ICL streams but also
the large radial extent of the halos of Virgo ellipticals. In
particular, M87's halo is traced beyond 150 kpc, where a variety of
signatures indicative of past accretion events can be seen. The
outermost regions of M87's halo are extremely boxy (Mihos \etal\ in
prep), a behavior reflected in the spatial distribution of its GC
system as well (Durrell \etal\ 2014). This combination of boxy isophotes
and low halo rotation (Romanowsky \etal\ 2012) hints at a major merger
event in M87's past, and indeed, both the GC and PNe systems around M87
show kinematic substructure (Romanowsky \etal\ 2012 and Longobardi
\etal\ 2015a respectively), suggesting the recent accretion of one or
more $\sim 10^{10}$ L\sun systems.

Signatures of past accretion are also found in other Virgo
ellipticals as well. Located south of the Virgo core, M49 has long been known to
have a dynamically complex halo, as traced by kinematic substructure in
its GC system (C\^ot\'e \etal\ 2003). The deep imaging in
Figure~\ref{virgo} reveals the cause: after subtraction of a smooth
isophotal model for M49, an extensive set of accretion shells
(Janowiecki \etal\ 2010, Arrigoni Battaia \etal\ 2011, Capaccioli \etal\
2015) can be seen, spanning $\sim$ 150 kpc in extent and containing
close to $10^9$ L\sun of light (Janowiecki \etal\ 2010). The shells are
morphologically similar to those formed during the radial accretion of a low mass
satellite, and may be linked to the tidally disturbed dwarf companion
VCC 1249 (Arrigoni Battaia \etal\ 2011). The shells are also distinctly
{\it redder} than M49's surrounding halo (Mihos \etal\ 2013), suggesting that
the accretion event is building up both the mass {\it and}
metallicity of M49's outer halo.

Figure~\ref{virgo} also illustrates the efficacy of the group environment
in driving ICL formation. Lying 5.3$^\circ$ to the SW of the Virgo core
(and $\sim$ 7 Mpc behind; Mei \etal\ 2007) is the infalling Virgo
W$^\prime$ group, with the massive elliptical NGC 4365 at its core. Our
deep imaging shows an extended, diffuse tidal tail emanating SW from the
galaxy (Bogd\'an \etal\ 2012; Mihos \etal\ in prep), and GC kinematics
clearly link the tail to an interaction with its companion NGC 4342
(Blom \etal\ 2014). The tail contains $\sim 1.5\times10^9$ L\sun\llap,
and a number of other streams are visible in NGC 4365's halo as well
(including the loop visible to the NE of the galaxy), all indicative of
cold tidal stripping in the group environment. Once the W$^\prime$ group
eventually falls into the main body of Virgo, this diffuse and extended
intragroup light will be easily mixed into Virgo's diffuse ICL.

Finally, the imaging contains a dramatic example of the complex
dynamical interplay between tidal stripping, ICL formation, and the
destruction and formation of cluster galaxy populations. Lying at the
center of the ``Tidal streams'' panel of Figure~\ref{virgo} is a large and
extremely dim ultra-diffuse galaxy; with a half light radius of 9.7 kpc
and central surface brightness $\mu_V=27.0$ it is the most extreme
ultradiffuse cluster galaxy yet discovered (Mihos \etal\ 2015). The
galaxy also sports a long tidal tail arcing $\sim$ 100 kpc to the north,
as well as a compact nucleus whose photometric properties are
well-matched to those of ultracompact dwarf galaxies (UCDs) found in
Virgo (\eg Zhang \etal\ 2015, Liu \etal\ 2015). In this object, we are
clearly seeing the tidal destruction of a low mass, nucleated galaxy
which is both feeding Virgo's ICL population and giving rise to a new
Virgo UCD.

To go beyond morphology and study the stellar populations in Virgo's
ICL in detail, a variety of tools are available. The colors of the streams around
M87 ($B-V=0.7-1.0$; Rudick \etal\ 2010) are well-matched to those of the
Virgo dE population and of M87's halo itself, suggesting M87's halo may
be built at least in part from low mass satellite accretion. HST imaging of
{\it discrete} RGB populations in Virgo intracluster fields shows the
ICL to be predominantly old and metal-poor ($t>10$ Gyr, [Fe/H] $\approx
-1$; Williams \etal\ 2007), but with an additional population of stars with
intermediate ages and higher metallicities ($t\approx 4-8$ Gyr, [Fe/H]
$\gtrsim -0.5$). These younger populations may arise either from stripped
star forming galaxies or from ICL formed {\it in-situ}. The inference
that stripping of late-type galaxies has contributed to the Virgo ICL is
also supported by the luminosity function of PNe in M87's outer halo,
which shows a ``dip'' characteristic of lower mass galaxies
with extended star formation histories (Longobardi \etal\ 2015b). The
diversity of stellar populations seen in Virgo's ICL almost certainly
reflects the diversity of processes that create diffuse light in
clusters.

\section{ICL Systematics: Challenges and Metrics}

While Virgo's proximity gives us a detailed view of intracluster stellar
populations, to gain a wider census of ICL in galaxy clusters we must
move beyond Virgo. Going to greater distances opens up the ability to
study ICL in a wider sample of clusters which span a range of mass,
dynamical state, and redshift, allowing us to connect ICL properties
with cluster evolution. This comes at a cost, however; beyond Virgo,
current generation telescopes cannot directly image intracluster stars,
and even studies of more luminous tracers such as PNe and GC become more
difficult. At higher redshifts, one becomes limited to broadband
imaging, where the strong cosmological $(1+z)^4$ surface brightness
dimming makes the already diffuse ICL even more difficult to observe.

Aside from these observational difficulties, a second major problem is
the ambiguous definition of intracluster light itself. Since much of the
ICL is formed via the mergers that build up the central BCG, there is
often no clear differentiation between the BCG halo and the extended ICL
-- the two components blend smoothly together (and indeed may not be
conceptually distinct components at all). In attempts to separate BCG
halos from extended ICL, a variety of photometric definitions have been
proposed, which typically adopt different functional forms (such as
multiple $r^{1/4}$ or Sersic profiles) for each component when fitting
the total profile (\eg Gonzalez \etal\ 2005, Krick \& Bernstein 2007,
Seigar \etal\ 2007). However, such definitions are very sensitive to the
functional forms adopted for the profiles. For example, M87's profile is
reasonably well-fit by either a single Sersic or a double $r^{1/4}$
model (Janowiecki \etal\ 2010); the former fit would imply little
additional ICL, while the latter fit puts equal light into the inner and
outer profiles. To avoid this ambiguity, alternate non-parametric
measures have also been employed to characterize the ICL luminosity,
defining the ICL as diffuse light fainter than some characteristic
surface brightness (\eg Feldmeier \etal\ 2004, Burke \etal\ 2015). While
these present systematic uncertainties of their own, simulations suggest
that thresholds of $\mu_V \gtrsim 26.5$ do a reasonable job of
separating out an extended and perhaps unrelaxed ICL component from the
central BCG light (Rudick \etal\ 2011, Cooper \etal\ 2015).

\begin{figure}[b]
\begin{center}
 \includegraphics[height=2in]{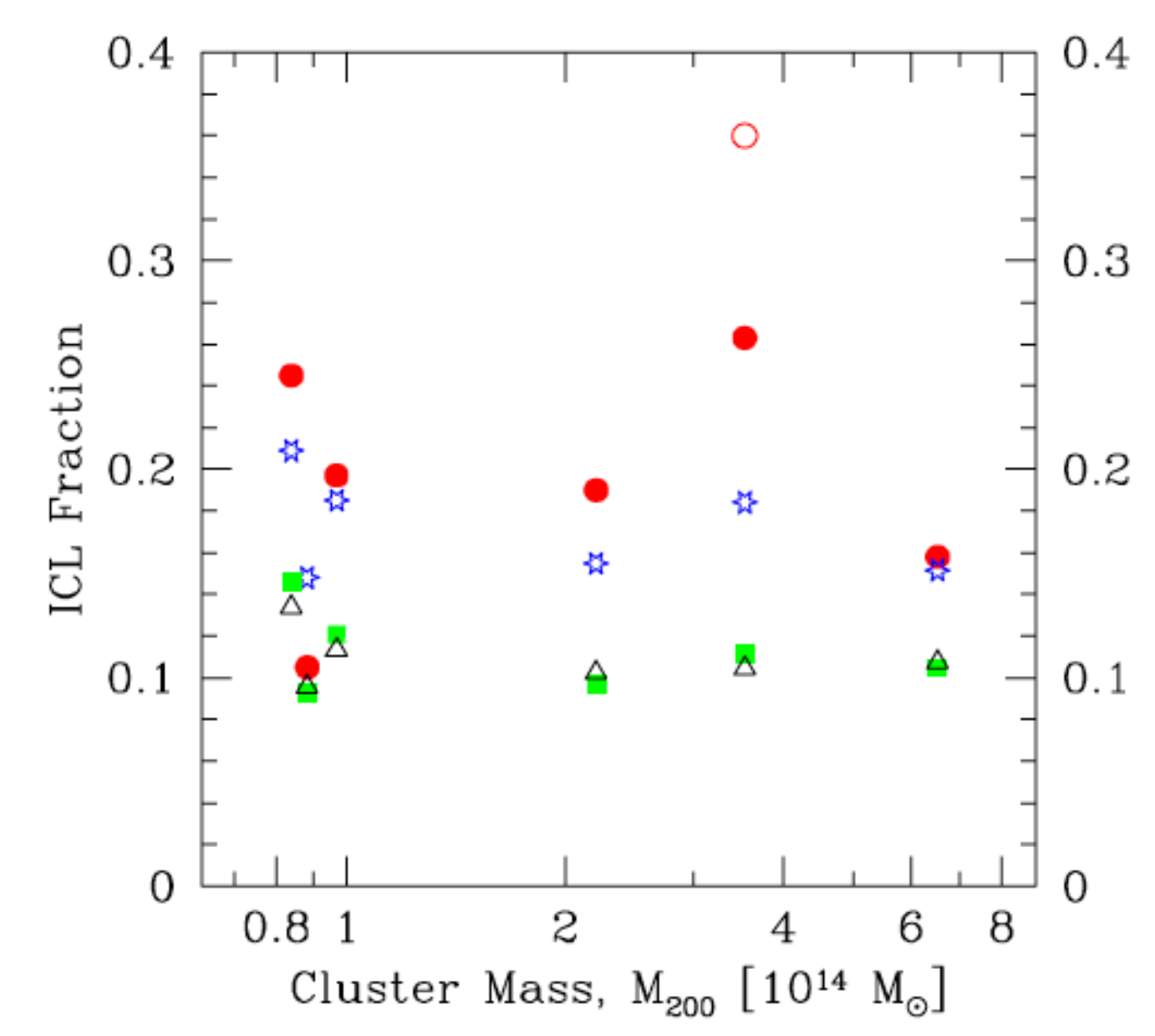}
 \caption{ICL fractions in simulated galaxy clusters, taken from Rudick \etal\  (2011). Five
 simulated clusters of varying mass are shown; at a given mass, the symbols show ICL fractions
 calculate for a single cluster using different ICL definitions: stars at low surface brightness ($\mu_V>26.5$, green squares), 
 stars in low density intracluster space now (open triangles) or ever (blue stars), stars unbound from galaxies (filled
 red circles), including stars kinematically separated from the central cD galaxy (open red circle).
 See Rudick \etal\  (2011) for details.}
   \label{ICLdef}
\end{center}
\end{figure}

Still other methods propose kinematic separation of the ICL from the
central galaxy light. Dolag \etal\ (2010) used simulated clusters to
show that separate kinematic populations exist in cluster cores,
well-characterized by distinct Maxwellian distributions. These kinematic
populations then separate out spatially into two Sersic-like profiles
plausibly identified as the BCG galaxy and the cluster ICL (perhaps
reflecting different accretion events as well; Cooper \etal\ 2015). And
indeed these definitions have some observational support. Longslit
spectroscopy of the BCG galaxy in Abell 2199 shows a velocity dispersion
profile that first falls with radius, then increases in the outer halo
to join smoothly onto the cluster velocity dispersion (Kelson \etal\
2002). Meanwhile in Virgo the velocities of the PNe around M87 show a
double Gaussian distribution (Longobardi \etal\ 2015b), suggesting
distinct BCG and ICL components. However, observational constraints make
accessing kinematic information for the ICL in distant clusters a
daunting task.

A comparison of these different metrics is shown in Figure~\ref{ICLdef}
(from Rudick \etal\ 2011), which shows that the inferred ICL fraction in
simulated clusters can vary by factors of 2$-$3 depending on the adopted
metric (see also Puchwein \etal\ 2010). Kinematic separation leads to
higher ICL fractions, as a significant amount of starlight found within
the BCG galaxy belongs to the high-velocity ICL component. In contrast,
density-based estimates yield systematically lower ICL fractions, as
material at high surface brightness is typically assigned to the cluster
galaxies independent of its kinematic properties.

Given both the ambiguity in defining the ICL and the observational
difficulties in studying it, attempts to characterize ICL in samples of
clusters spanning a range of mass and redshift have led to varying
results. An early compilation of results for local clusters by
Ciardullo \etal\ (2004) showed ICL fractions ranging from
$\sim 15-40$\%, with no clear dependence on cluster
velocity dispersion or Bautz-Morgan type. Recent imaging of more
distant clusters probes the connection between cluster evolution and
ICL more directly, but again yields mixed results. While
Guennou \etal\ (2012) find no strong difference between the ICL content
of clusters between at $z\sim0.5$ and today, Burke \etal\ (2015) find
rapid evolution in the ICL fraction of massive clusters over a similar
redshift range. Other studies of clusters at $z\sim 0.3-0.5$ yield ICL fractions
of 10$-$25\% (Presotto \etal\ 2014, Montes \& Trujillo 2014, Giallongo \etal\
2014), similar to $z=0$ results. However, these studies use different
ICL metrics and are limited to only a handful of clusters; clearly a large sample
of clusters with ICL fractions measured in a consistent manner is needed
to tackle the complex question of ICL evolution.

A similar story holds for recent attempts to constrain ICL stellar
populations as well. Using HST imaging of distant CLASH clusters, DeMaio
\etal\ (2015) infer moderately low metallicities ([Fe/H] $\sim -0.5$)
from the ICL colors, in contrast to the case of Abell 2744, where Montes
\& Trujillo (2014) use colors to argue for a dominant population of intermediate age
stars with solar metallicity. Meanwhile, spectroscopic population
synthesis studies show similarly diverse results. For example, in the
Hydra I cluster, Coccato \etal\ (2011) find old ICL populations with
sub-solar metallicities, while in the massive cluster RX J0054.0$-$2823,
Melnick \etal\ (2012) find similarly old but metal-rich ICL stars
([Fe/H] $\gtrsim 0$). However, while intriguing, all these studies are
subject to strong photometric biases, limited largely to the brightest
portions of the ICL which may not be representative of the ICL as a
whole and may also include substantial fraction of what would normally
be considered BCG light as well.

\section{Diffuse Light in Nearby Galaxy Groups}

The evolution shown in Figure~\ref{sim} argues that interactions in the
group environment should be particularly effective at stripping stars
from galaxies and redistributing them into the diffuse intragroup light,
an important precursor to the ICL in massive clusters. Curiously,
though, these arguments have not always been borne out observationally.
In the nearby Leo I group, searches for intragroup light using both PNe
(Castro-Rodriguez \etal\ 2003) and broadband imaging (Watkins \etal\
2014) have come up empty, particularly notable given that the system is
contains a large ($\sim$ 200 kpc) HI ring thought to be collisional in
origin (Michel-Dansac \etal\ 2010). Similarly, the M101 group also show
little sign of extended diffuse light (Mihos \etal\ 2013), despite the
tidal disturbances evident in M101 and its nearby companions. Even in
the clearly interacting M81/M82 group, early searches for orphaned
RGB stars (Durrell \etal\ 2004) and PNe (Feldmeier \etal\
2003) could only place upper limits on the intragroup light fraction
($\lesssim$ 2\%).

\begin{figure}[b]
\begin{center}
 \includegraphics[height=2in]{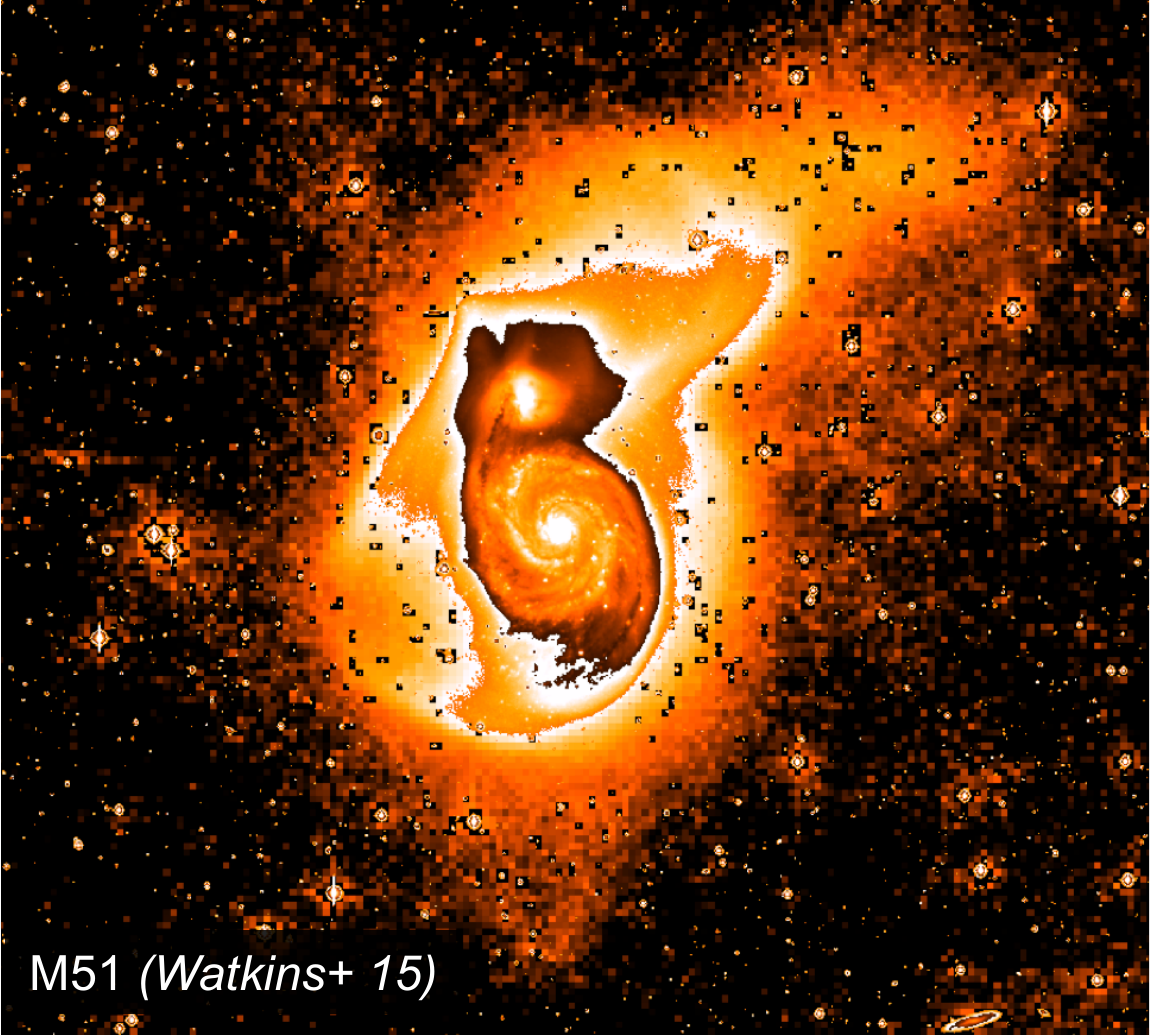}\includegraphics[height=2in]{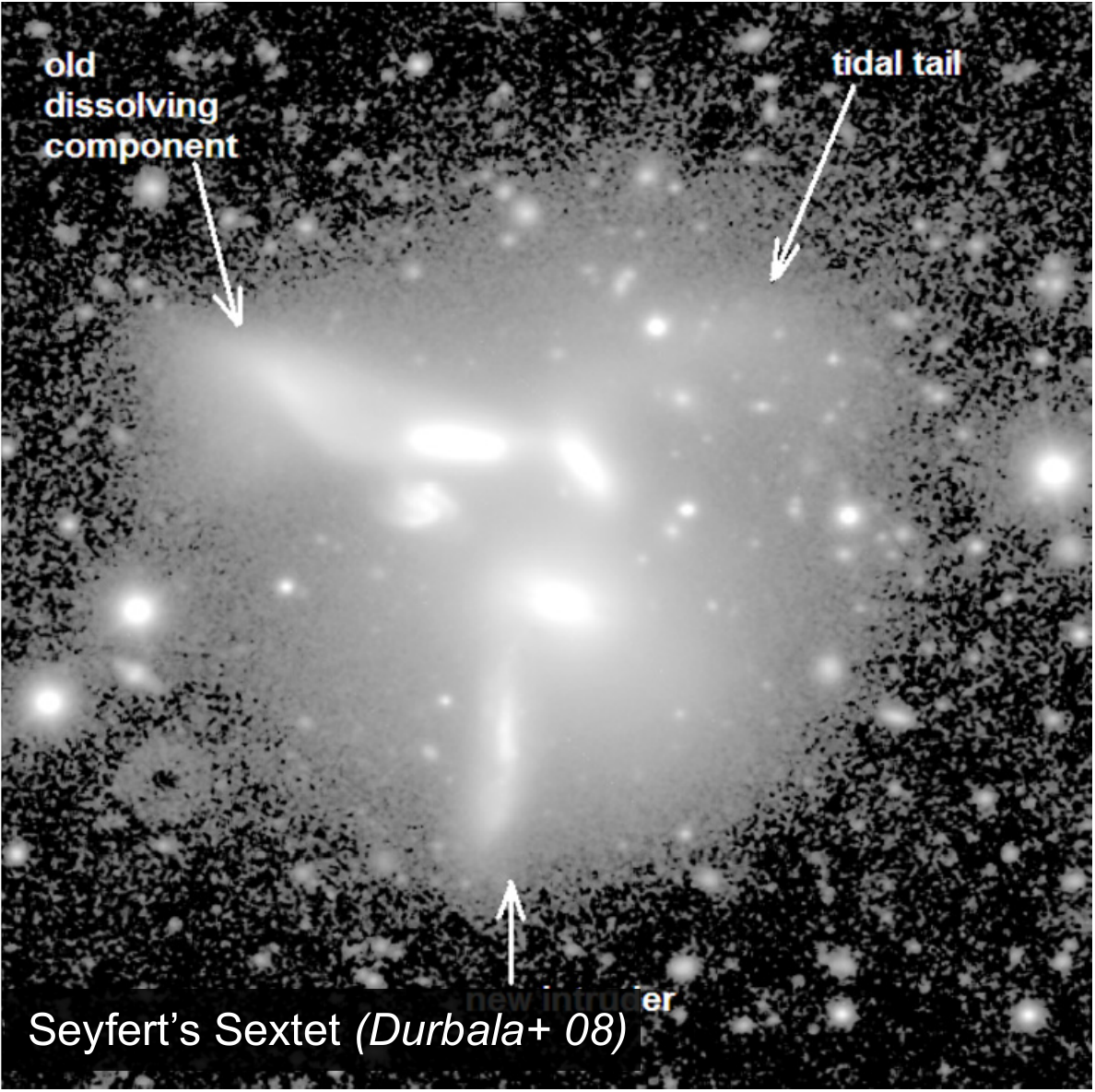}
 \end{center}
   \caption{Left: Deep imaging of M51 showing diffuse tidal debris extending
    to nearly 50 kpc (Watkins \etal\ 2015). Right: Diffuse light in Seyfert's Sextet 
   (Durbala \etal\ 2008).}
    \label{groups}
\end{figure}

In contrast, intragroup light is quite evident in dense, strongly
interacting groups. The ability for these strong interactions to expel
diffuse material to large distances is shown in recent deep imaging of
the M51 system by Watkins \etal\ (2015; Fig~\ref{groups}a), where
several extremely low surface brightness plumes extend nearly 50 kpc
from the center. Similarly, many compact groups are awash in diffuse
light (\eg Da Rocha \etal\ 2005, 2008), including the archetypal groups
Seyfert's Quintet (Mendes de Oliveira \etal\ 2001) and Seyfert's Sextet
(Durbala \etal\ 2008, Figure~\ref{groups}b). The contrast between the
copious diffuse light seen in these dense systems and the dearth of light
in loose groups is striking, arguing either that the tidal debris is
rapidly dispersed to even lower (undetectable) surface brightnesses, or
that close interactions in loose groups are relatively uncommon.

More recently, the ability to probe discrete stellar populations in
external galaxies provides a powerful new tool for studying
intragroup light. Probing stellar densities far below the capabilities
of wide-area surface photometry, these techniques are now revealing the
diffuse light contained even in loose groups. Deep imaging by Okamoto
\etal\ (2015) has uncovered the previously undetected and very extended stellar tidal debris
field in the M81 group, while imaging of M31 and M33 by the PAndAS team
(\eg Ibata \etal\ 2014) has mapped the myriad of tidal streams that
characterize Andromeda's extended stellar halo and trace its past interaction
with M33. While at low surface brightness and containing only a small
amount of the total light of their parent groups, the diffuse starlight
found in these studies of nearby loose groups represent the important
first step in building the intragroup and intracluster light in dense
galaxy environments.

\end{document}